\documentclass[11pt]{article}
\pdfoutput=1
\usepackage{amsmath,amssymb,amsthm,amsxtra,overpic,bbm,bm,epsfig,ulem,color,multirow}
\usepackage{afterpage}
\usepackage{multirow}
\usepackage{rotating}

\newcommand{\Dmq}{\Delta m^2}

\newcommand{\eVq}{\ensuremath{\text{eV}^2}}

\begin{document}

\begin{center}
{\Large \bf Particular textures of the minimal seesaw model}
\end{center}

\vspace{0.05cm}

\begin{center}
{\bf Zhen-Hua Zhao, Yan-Bin Sun\footnote{Corresponding author: sunyb@lnnu.edu.cn} and Shan-Shan Jiang} \\
{ Department of Physics, Liaoning Normal University, Dalian 116029, China }
\end{center}

\vspace{0.2cm}

\begin{abstract}
Following the Occam's razor principle, we explore the particular textures (featuring texture zeros and equalities) of the Dirac neutrino mass matrix in the minimal seesaw model (where only two right-handed neutrinos $N^{}_1$ and $N^{}_2$ are responsible for the generation of the light neutrino masses), in light of current neutrino experimental results and leptogenesis. The study will be performed for two particular patterns of the Majorana mass matrix $M^{}_{\rm R}$ for $N^{}_1$ and $N^{}_2$: (A) $M^{}_{\rm R}$ being diagonal $ \left( \begin{array}{cc}  M^{}_1 & 0 \cr
0 & M^{}_2 \cr  \end{array} \right)$; (B) $M^{}_{\rm R}$ being of the form $\left( \begin{array}{cc}  0 & M \cr
M & 0 \cr  \end{array} \right)$.
\end{abstract}

\newpage

\section{Introduction}

The observation of neutrino oscillations establishes that neutrinos are massive and the lepton flavors are mixed \cite{xing}. In the literature, the most popular and natural way of generating the small but non-zero neutrino masses is the type-I seesaw mechanism, in which three right-handed neutrinos are added into the SM and the lepton number conservation is violated by their Majorana mass terms \cite{seesaw}. The essence of the seesaw mechanism is to place the right-handed neutrinos at a scale much higher than the electroweak scale. Integrating them out naturally yields the small light neutrino masses (which are suppressed by the heavy right-handed neutrino masses)
\begin{eqnarray}
M^{}_{\nu} \simeq - M^{}_{\rm D} M^{-1}_{\rm R} M^{T}_{\rm D} \; ,
\label{1}
\end{eqnarray}
where $M^{}_{\rm D}$ is the $3 \times 3$ Dirac neutrino mass matrix connecting the left- and right-handed neutrinos and $M^{}_{\rm R}$ is the $3 \times 3$ Majorana mass matrix for the right-handed neutrinos. Then, in the basis where the flavor eigenstates of three charged leptons are identical with their mass eigenstates, the lepton flavor mixing matrix $U$ \cite{pmns} is to be identified as the unitary matrix for diagonalizing $M^{}_\nu$
\begin{eqnarray}
U^\dagger M^{}_\nu U^* =  {\rm Diag}(m^{}_1, m^{}_2, m^{}_3) \equiv D^{}_\nu   \;,
\label{2}
\end{eqnarray}
with $m^{}_i$ (for $i=1, 2, 3$) being the light neutrino masses.
In the standard parametrization, $U$ is expressed in terms of three mixing angles $\theta^{}_{ij}$ (for $ij=12, 13, 23$), one Dirac CP phase $\delta$ and two Majorana CP phases $\rho$ and $\sigma$ as
\begin{eqnarray}
U  =
\left( \begin{matrix}
c^{}_{12} c^{}_{13} & s^{}_{12} c^{}_{13} & s^{}_{13} e^{-{\rm i} \delta} \cr
-s^{}_{12} c^{}_{23} - c^{}_{12} s^{}_{23} s^{}_{13} e^{{\rm i} \delta}
& c^{}_{12} c^{}_{23} - s^{}_{12} s^{}_{23} s^{}_{13} e^{{\rm i} \delta}  & s^{}_{23} c^{}_{13} \cr
s^{}_{12} s^{}_{23} - c^{}_{12} c^{}_{23} s^{}_{13} e^{{\rm i} \delta}
& -c^{}_{12} s^{}_{23} - s^{}_{12} c^{}_{23} s^{}_{13} e^{{\rm i} \delta} & c^{}_{23}c^{}_{13}
\end{matrix} \right) \left( \begin{matrix}
e^{{\rm i}\rho} &  & \cr
& e^{{\rm i}\sigma}  & \cr
&  & 1
\end{matrix} \right) \;,
\label{3}
\end{eqnarray}
where the abbreviations $c^{}_{ij} = \cos \theta^{}_{ij}$ and $s^{}_{ij} = \sin \theta^{}_{ij}$ have been used.

There are totally six parameters governing the neutrino oscillation behaviors: $\theta^{}_{12}$, $\theta^{}_{13}$, $\theta^{}_{23}$, $\delta$ and two independent neutrino mass squared differences $\Delta m^2_{ij} \equiv m^2_i - m^2_j$ (for $ij =21, 31$). Thanks to the various neutrino oscillation experiments, these parameters have been determined to a good accuracy except that the result for $\delta$ is subject to a large uncertainty, and the sign of $\Delta m^2_{31}$ remains undetermined which allows for two possible neutrino mass orderings: the normal ordering (NO) $m^{}_1 < m^{}_2 < m^{}_3$ and inverted ordering (IO) $m^{}_3 < m^{}_1 < m^{}_2$. Several groups have performed a global analysis of the existing neutrino oscillation data to extract or constrain the values of these parameters and obtained mutually consistent results \cite{global,global2}. For concreteness, we will use the results in Ref.~\cite{global} (see Table~\ref{data}) as a representative in the subsequent numerical calculations. However, neutrino oscillations have nothing to do with the absolute neutrino mass scale and Majorana CP phases. In order to extract or constrain their values, one has to resort to some non-oscillatory experiments (e.g., the neutrino-less double beta decay experiments). Unfortunately, these experiments have not yet placed any lower constraint on the absolute neutrino mass scale, nor any constraint on the Majorana CP phases.

\begin{table}\centering
  \begin{footnotesize}
    \begin{tabular}{c|cc|cc}
     \hline\hline
      & \multicolumn{2}{c|}{Normal Ordering}
      & \multicolumn{2}{c}{Inverted Ordering }
      \\
      \cline{2-5}
      & bf $\pm 1\sigma$ & $3\sigma$ range
      & bf $\pm 1\sigma$ & $3\sigma$ range
      \\
      \cline{1-5}
      \rule{0pt}{4mm}\ignorespaces
       $\sin^2\theta_{12}$
      & $0.304_{-0.012}^{+0.013}$ & $0.269 \to 0.343$
      & $0.304_{-0.012}^{+0.013}$ & $0.269 \to 0.343$
      \\[1mm]
       $\sin^2\theta_{23}$
      & $0.570_{-0.024}^{+0.018}$ & $0.407 \to 0.618$
      & $0.575_{-0.021}^{+0.017}$ & $0.411 \to 0.621$
      \\[1mm]
       $\sin^2\theta_{13}$
      & $0.02221_{-0.00062}^{+0.00068}$ & $0.02034 \to 0.02430$
      & $0.02240_{-0.00062}^{+0.00062}$ & $0.02053 \to 0.02436$
      \\[1mm]
       $\delta/\pi$
      & $1.08_{-0.14}^{+0.28}$ & $0.59 \to 2.24$
      & $1.59_{-0.18}^{+0.15}$ & $1.07 \to 2.00$
      \\[3mm]
       $\dfrac{\Dmq_{21}}{10^{-5}~\eVq}$
      & $7.42_{-0.20}^{+0.21}$ & $6.82 \to 8.04$
      & $7.42_{-0.20}^{+0.21}$ & $6.82 \to 8.04$
      \\[3mm]
       $\dfrac{\Dmq_{31}}{10^{-3}~\eVq}$
      & $+2.514_{-0.027}^{+0.028}$ & $+2.431 \to +2.598$
      & $-2.423_{-0.028}^{+0.028}$ & $-2.509 \to -2.338$
      \\[2mm]
      \hline\hline
    \end{tabular}
  \end{footnotesize}
  \caption{The best-fit values, 1$\sigma$ errors and 3$\sigma$ ranges of six neutrino
oscillation parameters extracted from a global analysis of the existing
neutrino oscillation data \cite{global}. }
\label{data}
\end{table}

It is a great bonus that the seesaw model, via the leptogenesis mechanism \cite{yanagida,review}, also offers an attractive explanation for the baryon asymmetry of the Universe \cite{planck}
\begin{eqnarray}
Y^{}_{\rm B} \equiv \frac{n^{}_{\rm B}-n^{}_{\rm \bar B}}{s} = (8.67 \pm 0.15) \times 10^{-11}  \;,
\label{4}
\end{eqnarray}
where $n^{}_{\rm B}$ ($n^{}_{\rm \bar B}$) is the baryon (anti-baryon) number density and $s$ the entropy density. The leptogenesis mechanism proceeds in a way as follows: a lepton asymmetry $Y^{}_{\rm L} \equiv (n^{}_{\rm L}-n^{}_{\rm \bar L})/s$ is first generated from the out-of-equilibrium, lepton-number- and CP-violating decays of the right-handed neutrinos and then partially converted into the baryon asymmetry by means of the non-perturbative sphaleron process \cite{sphaleron}: $Y^{}_{\rm B} = -c Y^{}_{\rm L}$ with $c = 28/79$ in the SM \cite{relation}.

Although the information about the low-energy neutrino observables is completely encoded in $M^{}_\nu$, it is still meaningful for us to examine the possible structures of $M^{}_{\rm D}$ and $M^{}_{\rm R}$  for the following two reasons. (1) Motivated by that the observed lepton flavor mixing can be well approximated by a special mixing scheme (e.g., the tribimaximal mixing \cite{TB} or its trimaximal variations \cite{TM}), many attempts have been made to explore the possible flavor symmetry underlying the lepton sector \cite{nonabelian}. Being more fundamental than $M^{}_\nu$, a study on the possible structures of $M^{}_{\rm D}$ and $M^{}_{\rm R}$ may better help us reveal the underlying flavor symmetry. (2) For some high-energy processes such as leptogenesis, the structures of $M^{}_{\rm D}$ and $M^{}_{\rm R}$ will become relevant.

Unfortunately, $M^{}_{\rm D}$ itself alone consists of much more free parameters (15, after the removal of three phases by the rephasing of three left-handed neutrino fields) than the low-energy neutrino observables (9), making it difficult to infer the possible structures of $M^{}_{\rm D}$ and $M^{}_{\rm R}$ in light of current experimental results. This compels us to reduce the free parameters of $M^{}_{\rm D}$ and $M^{}_{\rm R}$.
A popular and natural way of doing so is to reduce the number of right-handed neutrinos. It turns out that the minimal number of the right-handed neutrinos necessary for reproducing the observed two non-zero neutrino mass squared differences is two. Furthermore, the minimal number of the right-handed neutrinos allowing for a successful leptogenesis is also two. Therefore, the seesaw model with only two right-handed neutrinos is the minimal one \cite{mss1,mss2,mss3}. In the minimal seesaw model, $M^{}_{\rm D}$ is reduced to a $3\times2$ matrix
\begin{eqnarray}
M^{}_{\rm D} = \left( \begin{array}{cc}  (M^{}_{\rm D})^{}_{e1} & (M^{}_{\rm D})^{}_{e2} \cr
(M^{}_{\rm D})^{}_{\mu 1} & (M^{}_{\rm D})^{}_{\mu 2} \cr (M^{}_{\rm D})^{}_{\tau 1} & (M^{}_{\rm D})^{}_{\tau 2} \end{array} \right) = \left( \begin{array}{cc}  a^{}_{1} & b^{}_{1} \cr
a^{}_{2} & b^{}_{2} \cr a^{}_{3} & b^{}_{3} \end{array} \right) \;,
\label{5}
\end{eqnarray}
where the symbols $(M^{}_{\rm D})^{}_{\alpha i}$ (for $\alpha =e, \mu, \tau$ and $i=1, 2$) and $a^{}_i$ and $b^{}_i$ (for $i=1, 2, 3$) will be used interchangeably in the following discussions. And $M^{}_{\rm R}$ becomes a $2 \times 2$ matrix. It is straightforward to verify that now ${\rm det}(M^{}_\nu) =0$ holds, implying that one of the light neutrino masses ($m^{}_1$ in the NO case or $m^{}_3$ in the IO case) necessarily vanishes. As an immediate consequence, one Majorana CP phase will become physically irrelevant, leaving us with only one effective Majorana CP phase ($\sigma$).
However, even in the minimal seesaw model, $M^{}_{\rm D}$ itself alone still consists of more free parameters (9, after the removal of three phases by the rephasing of three left-handed neutrino fields) than the low-energy neutrino observables (7). One can further reduce the free parameters by imposing texture zeros (which are usually tied to Abelian flavor symmetries \cite{abelian}) and equalities (which are usually tied to non-Abelian flavor symmetries \cite{nonabelian}) on $M^{}_{\rm D}$ and $M^{}_{\rm R}$. In this connection, the Frampton-Glashow-Yanagida (FGY) model \cite{mss1} serves as a unique example, where $M^{}_{\rm D}$ assumes two vanishing entries (in the scenario of $M^{}_{\rm R}$ being diagonal) and thus consists of fewer free parameters than the low-energy neutrino observables instead, leading to two predictions for them.

In this paper, for the minimal seesaw model, following the Occam's razor principle, we will explore the particular textures (featuring texture zeros and equalities) of $M^{}_{\rm D}$ in light of current experimental results and leptogenesis, for two particular patterns of $M^{}_{\rm R}$: (A) $M^{}_{\rm R}$ being diagonal $M^{}_{\rm R} ={\rm diag}(M^{}_1, M^{}_2) \equiv D^{}_{\rm R}$ (in section~2); (B) $M^{}_{\rm R}$ being of the form (in section~3)
\begin{eqnarray}
M^{}_{\rm R} = \left( \begin{array}{cc}  0 & M \cr
M & 0 \cr  \end{array} \right) \;.
\label{6}
\end{eqnarray}

\section{In the scenario of $M^{}_{\rm R}$ being diagonal}

In this section, we perform the study for Scenario A (i.e., $M^{}_{\rm R}$ being diagonal). Compelled by the fact that in the NO case the FGY model (i.e., two-zero textures of $M^{}_{\rm D}$) has been ruled out by current experimental results \cite{FGY}, we (in the NO case, accordingly) turn to
{\footnote{In Ref.~\cite{Ml}, the authors have studied the possibility of modifying the FGY model by considering the corrections from a non-diagonal charged lepton mass matrix.} }
the one-zero textures of $M^{}_{\rm D}$ \cite{onezero} and further examine if some equalities among the non-vanishing entries can also hold \cite{hybrid}. We first figure out the phenomenologically viable particular textures of $M^{}_{\rm D}$ in section~2.1 and then study their implications for leptogenesis in section~2.2.

\subsection{Particular textures of $M^{}_{\rm D}$}

\begin{figure}
\centering
\includegraphics[width=6in]{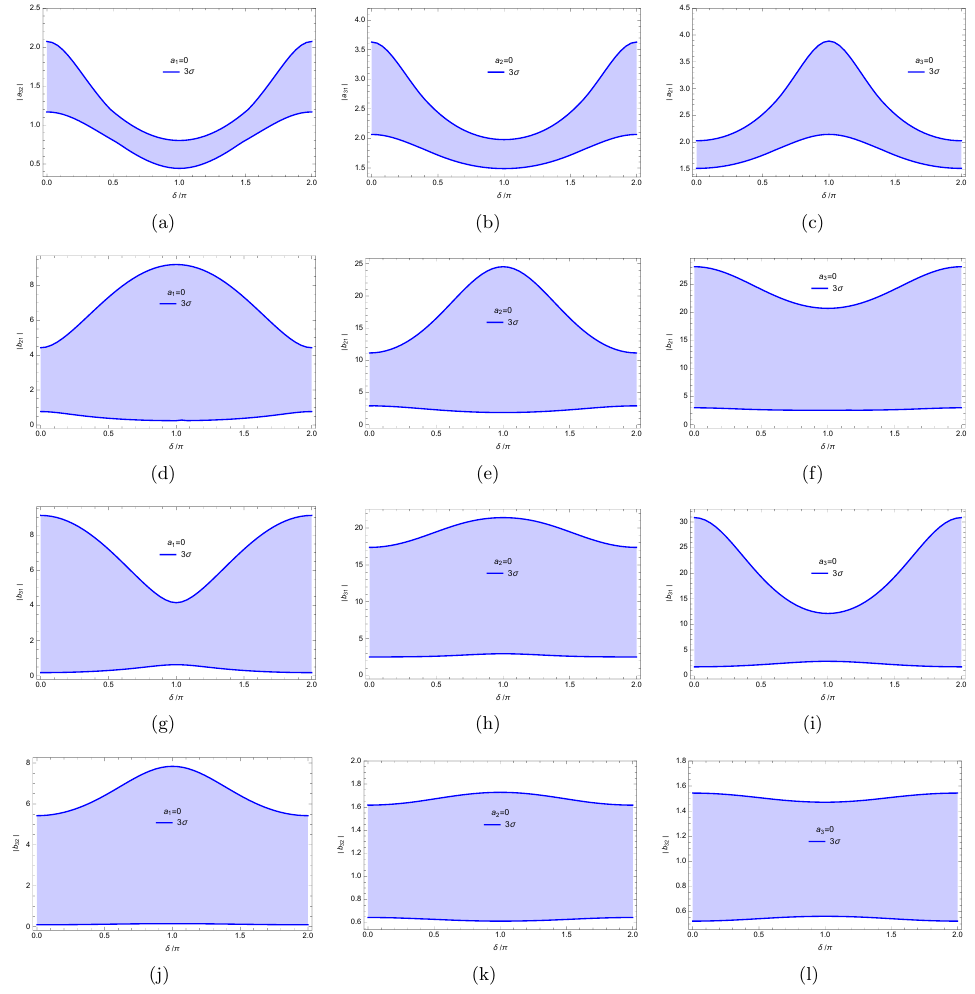}
\caption{ In Scenario A and the cases of $a^{}_i =0$, the allowed ranges of $|a^{}_{ij}| = |a^{}_i|/|a^{}_j|$ and $|b^{}_{ij}| = |b^{}_i|/|b^{}_j|$ as functions of $\delta$. In obtaining these results, $\sigma$ is allowed to vary in the whole range of $[0, \pi]$, while the other parameters in their $3\sigma$ ranges. }
\label{ratio}
\end{figure}

For our purpose, it will be convenient to make use of the Casas-Ibarra parametrization for $M^{}_{\rm D}$ \cite{CI}:
in the basis of $M^{}_{\rm R}$ being diagonal $D^{}_{\rm R}$, with the help of Eqs.~(\ref{1}, \ref{2}), it is easy to verify that $M^{}_{\rm D}$ can be expressed as \begin{eqnarray}
M^{}_{\rm D} = {\rm i} U \sqrt{D^{}_\nu} R \sqrt{D^{}_{\rm R}}  \;,
\label{2.1.1}
\end{eqnarray}
where $R$ can be parameterized as
\begin{eqnarray}
{\rm NO}: \hspace{1cm} R = \left( \begin{array}{cc} 0 & 0 \cr \cos{z} & -  \sin{z} \cr \sin{z} & \cos{z} \end{array} \right) \;;  \hspace{1cm}
{\rm IO}: \hspace{1cm} R = \left( \begin{array}{cc} \cos{z} & - \sin{z} \cr \sin{z} &  \cos{z} \cr 0 & 0 \end{array} \right) \;,
\label{2.1.2}
\end{eqnarray}
with $z$ being a complex parameter satisfying $\cos^2 z + \sin^2 z =1$. In such a parametrization, the two degrees of freedom $M^{}_{\rm D}$ contains more than the low-energy neutrino observables are encoded in $z$. In the NO case under consideration, the entries of $M^{}_{\rm D}$ read
\begin{eqnarray}
\left(M^{}_{\rm D}\right)^{}_{\alpha 1} = +{\rm i}\sqrt{M^{}_1} \left( U^{}_{\alpha 2} \sqrt{m^{}_2} \cos z +  U^{}_{\alpha 3} \sqrt{m^{}_3} \sin z \right) \;, \nonumber \\
\left(M^{}_{\rm D}\right)^{}_{\alpha 2} = - {\rm i} \sqrt{M^{}_2} \left( U^{}_{\alpha 2} \sqrt{m^{}_2} \sin z -  U^{}_{\alpha 3} \sqrt{m^{}_3} \cos z \right) \;.
\label{2.1.3}
\end{eqnarray}

\begin{figure*}
\centering
\includegraphics[width=6in]{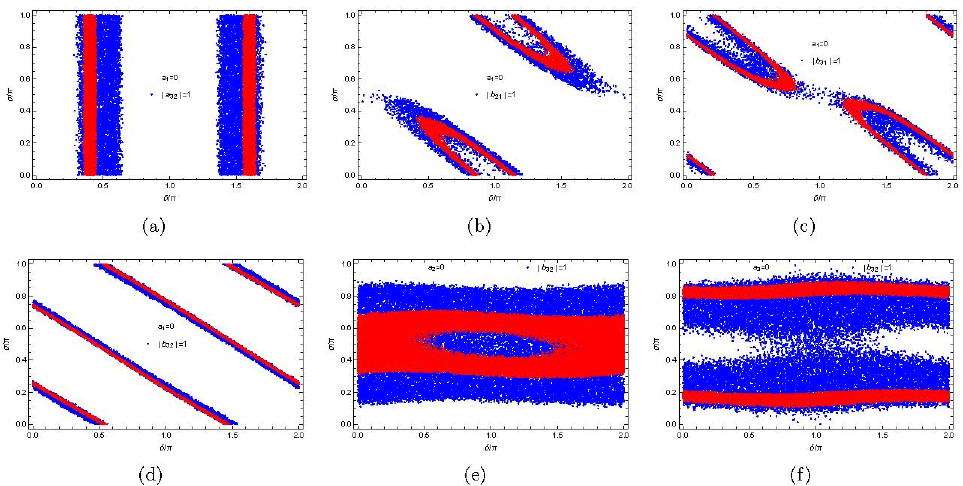}
\caption{ In Scenario A and the cases of $a^{}_i =0$, the allowed values of $\sigma$ versus $\delta$ for the further impositions of $a^{}_{i} =a^{}_{j}$ and $b^{}_{i} = b^{}_{j}$, at the $1\sigma$ (red) and $3\sigma$ (blue) levels. (For interpretation of the colors in the figure(s), the reader is referred to the web version of this article.)}
\label{sigdel}
\end{figure*}

Now we are ready to study the one-zero textures of $M^{}_{\rm D}$. We just need to consider the cases of $(M^{}_{\rm D})^{}_{\alpha 1} =0$, while the corresponding cases of $(M^{}_{\rm D})^{}_{\alpha 2} =0$ will lead to the same low-energy consequences. This is because $M^{}_\nu$ keeps invariant under the simultaneous interchanges $M^{}_{1} \leftrightarrow M^{}_2$ and two columns of $M^{}_{\rm D}$. From Eq.~(\ref{2.1.3}) it is easy to see that the imposition of $a^{}_i =0$  will lead to a determination of $z$ in terms of the low-energy neutrino observables
\begin{eqnarray}
\tan z = - \frac{ U^{}_{\alpha 2} \sqrt{m^{}_2} }{ U^{}_{\alpha 3} \sqrt{m^{}_3} } \;.
\label{2.1.4}
\end{eqnarray}
With the help of this result, one can figure out the ratios among the non-vanishing entries to examine if some of them can take the value of 1. Here we just consider the ratios among the non-vanishing entries residing in the same column, which are independent of the unknown right-handed neutrino masses.

\begin{table}[t]
\centering
\begin{tabular}{c|cccccc}
\hline\hline
&$|a^{}_{21}|$ & $|a^{}_{31}|$ & $|a^{}_{32}|$ & $|b^{}_{21}|$ & $|b^{}_{31}|$ & $|b^{}_{32}|$\ \\ \cline{1-7}
$a^{}_1 =0$  & $-$         &  $-$         & 0.44$-$2.07  & 0.24$-$9.20  & 0.18$-$9.12  & 0.10$-$7.84 \\
$a^{}_2 =0$  & 0           &  1.49$-$3.63 & $-$          & 1.89$-$24.54 & 2.52$-$21.41 & 0.61$-$1.73 \\
$a^{}_3 =0$  & 1.51$-$3.89 &  0           & 0            & 2.57$-$28.18 & 1.71$-$30.87 & 0.52$-$1.54 \\
\hline\hline
\end{tabular}
\caption{In Scenario A and the cases of $a^{}_i =0$, the allowed ranges of $|a^{}_{ij}|$ and $|b^{}_{ij}|$ at the $3\sigma$ level. }
\label{range}
\end{table}

Let us first consider the case of $a^{}_1 =0$. In Fig.~\ref{ratio} we plot the allowed ranges of $|a^{}_{32}| = |a^{}_3|/|a^{}_2|$, $|b^{}_{21}| = |b^{}_2|/|b^{}_1|$, $|b^{}_{31}| = |b^{}_3|/|b^{}_1|$ and $|b^{}_{32}| = |b^{}_3|/|b^{}_2|$ as functions of $\delta$, for which the numerical values are listed in Table~\ref{range}. One can see that $|b^{}_{21}|$, $|b^{}_{31}|$ and $|b^{}_{32}|$ can be very small, indicating that the FGY model with $a^{}_1 = b^{}_2 =0$ or $a^{}_1 = b^{}_3 =0$ can still hold as a good approximation \cite{approx}.
And all these quantities have chance to take the value of 1. By utilizing the freedom of the rephasing of three left-handed neutrino fields, $a^{}_{2} =a^{}_{3}$ (or $b^{}_{i} =b^{}_{j}$) can always be achieved from $|a^{}_{2}| =|a^{}_{3}|$ (or $|b^{}_{i}| =|b^{}_{j}|$). The further imposition of $a^{}_{2} =a^{}_{3}$ (or $b^{}_{i} =b^{}_{j}$) will lead us to a constraint on the low-energy neutrino observables
\begin{eqnarray}
(M^{}_{\rm D})^{}_{\mu 1} = (M^{}_{\rm D})^{}_{\tau 1}: \hspace{0.5cm}  |U^{}_{e 2} U^{}_{\mu 3} - U^{}_{e 3} U^{}_{\mu 2}| =  |U^{}_{e 2} U^{}_{\tau 3} - U^{}_{e 3} U^{}_{\tau 2} | \;; \nonumber \\
(M^{}_{\rm D})^{}_{\alpha 2} = (M^{}_{\rm D})^{}_{\beta 2}: \hspace{0.5cm}  | m^{}_2 U^{}_{e 2}  U^{}_{\alpha 2} + m^{}_3 U^{}_{e 3} U^{}_{\alpha 3} | = | m^{}_2 U^{}_{e 2}  U^{}_{\beta 2} + m^{}_3 U^{}_{e 3} U^{}_{\beta 3} | \;.
\label{2.1.5}
\end{eqnarray}
Taking account these constraints, in Fig.~\ref{sigdel} we plot the allowed values of $\sigma$ as functions of $\delta$ for the further impositions of $a^{}_{2} =a^{}_{3}$ (a), $b^{}_{1} =b^{}_{2}$ (b), $b^{}_{1} =b^{}_{3}$ (c) and $b^{}_{2} =b^{}_{3}$ (d). The results show that in the $a^{}_{2} =a^{}_{3}$ case $\delta$ is constrained to be around $\pm \pi/2$ while $\sigma$ is subject to no constraint (which can be easily verified by using Eq.~(\ref{2.1.5})). In the $b^{}_{i} =b^{}_{j}$ cases, there is a strong correlation between $\sigma$ and $\delta$, which can help us infer the former when the latter is determined by future neutrino oscillation experiments. Finally, we examine if two of $a^{}_{2} =a^{}_{3}$ and $b^{}_{i} =b^{}_{j}$ can hold simultaneously. It should be noted that $a^{}_{2} =a^{}_{3}$ and $b^{}_{2} =b^{}_{3}$ can never hold simultaneously, because otherwise $M^{}_{\rm D}$ would acquire a $\mu$-$\tau$ interchange symmetry \cite{mutau1,mutau2} which gives the unrealistic result $\theta^{}_{13} =0$. Our numerical calculations show that $a^{}_{2} =a^{}_{3}$ and $b^{}_{1} =b^{}_{2}$ ($b^{}_{1} =b^{}_{3}$) have chance to hold simultaneously. The consequences of these two cases for the low-energy neutrino observables (see Table~\ref{prediction}) are obtained by minimizing the $\chi^2$ function
\begin{eqnarray}
\chi^2 = \sum_i \left( \frac{ {\mathcal O}^{}_{i} - \overline {\mathcal O}^{}_{i} }{ \sigma^{}_{i } } \right)^2 \; ,
\label{2.1.6}
\end{eqnarray}
where the sum is over three mixing angles, two neutrino mass squared differences and $\delta$, and ${\mathcal O}^{}_i$, $\overline {\mathcal O}^{}_i$ and $\sigma^{}_{i }$ denote their predicted values, best-fit values and $1\sigma$ errors, respectively. For both cases, $\delta \simeq 3/2 \pi$ and $\theta^{}_{23} > 45^\circ$ are favored, in good agreement with current experimental results. It is worth pointing out that in the interesting littlest seesaw model \cite{LS} $M^{}_{\rm D}$ just has a texture featuring $a^{}_1 =0$ \& $a^{}_{2} =a^{}_{3}$ \& $b^{}_{1} =b^{}_{3}$.

\begin{table}[t]
\centering
\begin{tabular}{c|ccccccccc}
\hline\hline
conditions & $\delta/\pi$ & $\sigma/\pi$ & $\Delta m^2_{21}$ & $\Delta m^2_{31}$  & $s^{2}_{12}$  & $ s^{2}_{23} $ & $s^{2}_{13}$ & $\chi^2_{\rm min}$ \ \\ \cline{1-9}
$a^{}_{1} =0$ \& $a^{}_{2} =a^{}_{3}$ \& $b^{}_{1} =b^{}_{2}$  & 1.56 &  0.67 & 7.46  & 2.515  & 0.307  & 0.546 & 0.02211 & 12   \\
$a^{}_{1} =0$ \& $a^{}_{2} =a^{}_{3}$ \& $b^{}_{1} =b^{}_{3}$ & 1.58 &  0.35 & 7.42  &  2.517  & 0.303  & 0.560 & 0.02221 & 12 \\

\hline\hline
\end{tabular}
\caption{In Scenario A and the case of $a^{}_1 =0$, the predictions for the low-energy neutrino observables from the further impositions of $a^{}_{2} =a^{}_{3}$ and $b^{}_{1} =b^{}_{2}$ ($b^{}_{1} =b^{}_{3}$). The units of $\Delta m^2_{21}$ and $\Delta m^2_{31}$ are $10^{-5}$ eV$^2$ and $10^{-3}$ eV$^2$, respectively. }
\label{prediction}
\end{table}

Then, we consider the case of $a^{}_2 =0$, for which the allowed ranges of $|a^{}_{31}| = |a^{}_3|/|a^{}_1|$, $|b^{}_{21}|$, $|b^{}_{31}|$ and $|b^{}_{32}|$ are also listed in Table~\ref{range}: $|b^{}_{21}|$ and $|b^{}_{31}|$ can be very large, indicating that the FGY model with $a^{}_2 = b^{}_1 =0$ can still hold as a good approximation.
On the other hand, only $|b^{}_{32}|$ (equivalently $b^{}_{32}$) has chance to take the value of 1. The further imposition of $b^{}_{2} =b^{}_{3}$ will lead us to a constraint on the low-energy neutrino observables like Eq.~(\ref{2.1.5}), by which the allowed values of $\sigma$ versus $\delta$ in Fig.~\ref{sigdel}(e).
Finally, we point out that the results for the case of $a^{}_3 =0$ are similar to those in the case of $a^{}_2 =0$. This reflects the fact that the neutrino sector possesses an approximate $\mu$-$\tau$ flavor symmetry \cite{mutau3}.

To summarize, in the scenario of $M^{}_{\rm R}$ being diagonal, for the NO case, the phenomenologically viable particular textures of $M^{}_{\rm D}$ are as follows
\begin{eqnarray}
\left( \begin{array}{cc}  0 & \times \cr
\Box & \times \cr \Box & \times \end{array} \right) \;, \hspace{1cm}
\left( \begin{array}{cc}  0 &  \Box \cr
\times & \Box \cr \times & \times \end{array} \right) \;, \hspace{1cm}
\left( \begin{array}{cc}  0 &  \Box \cr
\times & \times \cr \times & \Box \end{array} \right) \;, \nonumber \\
\left( \begin{array}{cc}  0 &  \times \cr
\times & \Box  \cr \times & \Box \end{array} \right) \;, \hspace{1cm}
\left( \begin{array}{cc}  \times & \times \cr
0 & \Box \cr  \times & \Box \end{array} \right) \;, \hspace{1cm}
\left( \begin{array}{cc}  \times & \times \cr
\times  & \Box \cr  0 & \Box \end{array} \right) \;,
\label{2.1.7}
\end{eqnarray}
where the $\Box$ ($\times$) symbol is used to mark the equal (unconstrained) entries. Furthermore, the following more restricted textures of $M^{}_{\rm D}$ can also be consistent with the experimental results
\begin{eqnarray}
\left( \begin{array}{cc}  0 & \diamondsuit \cr
\Box & \diamondsuit \cr \Box & \times \end{array} \right) \;, \hspace{1cm}
\left( \begin{array}{cc}  0 &  \diamondsuit \cr
\Box &  \times \cr \Box & \diamondsuit \end{array} \right) \;,
\label{2.1.8}
\end{eqnarray}
where the $\diamondsuit$ symbol is used to mark another pair of equal entries. Finally, the textures of $M^{}_\nu$ that correspond to these textures of $M^{}_{\rm D}$ are listed in Table~\ref{texture}.

\begin{table*}
\centering
\begin{tabular}{c |c |c |c }
\hline\hline
&&{Scenario A} & {Scenario B} \\
\hline
\multirow{6}{*}{$a_{1}=0$}
& {$a_{2}=a_{3}$} & $-$  & $-$    \\
\cline{2-4}
& {$b_{1}=b_{2}$} & {$M_{ee}$=$M_{e\mu}$} & $-$  \\
\cline{2-4}
& {$b_{1}=b_{3}$} & {$M_{ee}$=$M_{e\tau}$} & $-$ \\
\cline{2-4}
& {$b_{2}=b_{3}$} & {$M_{e\mu}$=$M_{e\tau}$} & $-$  \\
\cline{2-4}
& {$a_{2}=a_{3}$ \& $b_{1}=b_{2}$} & {$M_{ee}$=$M_{e\mu}$ \& $M_{\mu\mu}$-$M_{e\mu}$=$M_{\mu\tau}$-$M_{e\tau}$} & $-$  \\
\cline{2-4}
& {$a_{2}=a_{3}$ \& $b_{1}=b_{3}$} & {$M_{ee}$=$M_{e\tau}$ \& $M_{\mu\tau}$-$M_{e\mu}$=$M_{\tau\tau}$-$M_{e\tau}$} & $-$  \\
\hline

\multirow{6}{*}{$a_{2}=0$}
& {$a_{i}=b_{j}$} & $-$ & {$M_{\mu\mu}=0$}\\
\cline{2-4}
& {$b_{2}=b_{3}$} & {$M_{\mu\mu}$=$M_{\mu\tau}$} & {$M_{\mu\mu}=0$ \& $M_{\tau\tau}$=$2 M_{\mu\tau}$}\\
\cline{2-4}
& {$a_{1}=b_{2}=b_{3}$} &$-$& {$M_{\mu\mu}=0$ \& $M_{\tau\tau}$=$2 M_{\mu\tau}$} \\
\cline{2-4}
& {$a_{3}=b_{2}=b_{3}$} &$-$& {$M_{\mu\mu}=0$ \& $M_{\tau\tau}$=$2 M_{\mu\tau}$} \\
\cline{2-4}
& {$a_{3}=b_{1}$ \& $b_{2}=b_{3}$} &$-$& {$M_{\mu\mu}=0$ \& $M_{\tau\tau}$=$2 M_{\mu\tau}$} \\
\cline{2-4}
& {$a_{1}=b_{1}$ \& $b_{2}=b_{3}$} &$-$& {$M_{\mu\mu}=0$ \& $M_{\tau\tau}$=$2 M_{\mu\tau}$} \\
\hline\hline
\end{tabular}
\caption{ In Scenario A and B, the textures of $M^{}_\nu$ (with $M^{}_{\alpha \beta}$ denoting its $\alpha \beta$ entry) that correspond to the texture of $M^{}_{\rm D}$ with $a^{}_i =0$ and the further impositions of some equalities among the non-vanishing entries. The results for the case of $a^{}_3 =0$ can be obtained from those for the case of $a^{}_2 =0$ by making the interchange $\mu \leftrightarrow \tau$. }
\label{texture}
\end{table*}

\subsection{Implications for leptogenesis}

In this subsection, we study the implications of the particular textures of $M^{}_{\rm D}$ in Eqs.~(\ref{2.1.7}, \ref{2.1.8}) for leptogenesis. Here we consider the scenario that there is a hierarchy between $M^{}_1$ and $M^{}_2$. For a hierarchical right-handed neutrino mass spectrum, the contribution to leptogenesis mainly comes from the lighter right-handed neutrino $N^{}_i$ (i.e., $i =1$ for $M^{}_1 < M^{}_2$ or $i =2$ for $M^{}_2 < M^{}_1$)\footnote{For some exceptional scenarios, see Refs.~\cite{N2}.}.

According to the temperature where leptogenesis takes place (approximately the mass of the lighter right-handed neutrino $M^{}_i$), there are several distinct leptogenesis regimes \cite{flavor}. (1) Unflavored regime: in the temperature range above $10^{12}$ GeV, the charged-lepton Yukawa ($y^{}_\alpha$) interactions have not yet entered thermal equilibrium, so the three lepton flavors are indistinguishable and to be treated universally. (2) Two-flavor regime: in the temperature range between $10^{12}$ GeV and $10^{9}$ GeV, the $y^{}_{\tau}$ related interactions enter thermal equilibrium but the $y^{}_{e}$ and $y^{}_{\mu}$ related interactions not, making the $\tau$ flavor distinguishable from the $e$ and $\mu$ flavors which remain indistinguishable. In this regime, the $\tau$ flavor should be treated separately from a coherent superposition of the $e$ and $\mu$ flavors. (3) Three-flavor regime: in the temperature range below $10^{9}$ GeV, the $y^{}_{\mu}$ related interactions also enter thermal equilibrium, making all the three lepton flavors distinguishable. In this regime, all the three lepton flavors should be treated separately.

In the two-flavor regime which is relevant for our study in this section, the $N^{}_i$-generated baryon asymmetry receives two contributions as follows \cite{flavor}
\begin{eqnarray}
Y^{}_{\rm B}
=  - c r \left[ \varepsilon^{}_{i \tau} \kappa \left(\frac{390}{589} \widetilde m^{}_{i \tau} \right) + \varepsilon^{}_{i \gamma} \kappa \left(\frac{417}{589} \widetilde m^{}_{i \gamma} \right) \right] \;,
\label{2.2.1}
\end{eqnarray}
with $\varepsilon^{}_{i \gamma} = \varepsilon^{}_{i e} + \varepsilon^{}_{i \mu}$ and $\widetilde m^{}_{i \gamma} = \widetilde m^{}_{i e} + \widetilde m^{}_{i \mu}$.
As mentioned above, $c= 28/79$ describes the transition efficiency from $Y^{}_{\rm L}$ to $Y^{}_{\rm B}$. $r \simeq 4 \times 10^{-3}$ is the ratio of the $N^{}_i$ number density to the entropy density at the temperature above $M^{}_i$. $\varepsilon^{}_{i \alpha}$ are the CP asymmetries for the decay processes of $N^{}_i$ \cite{yanagida,cp}
\begin{eqnarray}
\varepsilon^{}_{i \alpha} & \equiv & \frac{  \Gamma(N^{}_i \to L^{}_\alpha + H) - \Gamma(N^{}_i \to \overline{L^{}_\alpha} + \overline{H}) }{ \sum^{}_\alpha \left[\Gamma(N^{}_i \to L^{}_\alpha + H) + \Gamma(N^{}_i \to \overline{L^{}_\alpha} + \overline{H} ) \right] }  \nonumber \\
& = & \frac{1}{8\pi (M^\dagger_{\rm D}
M^{}_{\rm D})^{}_{ii} v^2} \left\{ {\rm Im}\left[(M^*_{\rm D})^{}_{\alpha i} (M^{}_{\rm D})^{}_{\alpha j}
(M^\dagger_{\rm D} M^{}_{\rm D})^{}_{ij}\right] {\cal F} \left( \frac{M^2_j}{M^2_i} \right) \right. \nonumber \\
&  &
+ \left. {\rm Im}\left[(M^*_{\rm D})^{}_{\alpha i} (M^{}_{\rm D})^{}_{\alpha j} (M^\dagger_{\rm D} M^{}_{\rm D})^*_{ij}\right] {\cal G}  \left( \frac{M^2_j}{M^2_i} \right) \right\} \; ,
\label{2.2.2}
\end{eqnarray}
where $v =174$ GeV is the Higgs vacuum expectation value, $j \neq i$, ${\cal F}(x) = \sqrt{x} \{(2-x)/(1-x)+ (1+x) \ln [x/(1+x)] \}$ and ${\cal G}(x) = 1/(1-x)$. In the Casas-Ibarra parametrization, $\varepsilon^{}_{1 \alpha}$ and $\varepsilon^{}_{2 \alpha}$ are explicitly expressed as
\begin{eqnarray}
\varepsilon^{}_{1\alpha} & \simeq & \frac{M^{}_2 }{8 \pi v^2 \left(m^{}_2 \left|\cos z\right|^2+ m^{}_3 \left|\sin z\right|^2\right)} \left[ A {\cal F} \left( \frac{M^2_2}{M^2_1} \right)  + B {\cal G} \left( \frac{M^2_2}{M^2_1} \right) \right] \; , \nonumber \\
\varepsilon^{}_{2 \alpha} & \simeq & - \frac{M^{}_1 }{8 \pi v^2 \left(m^{}_2 \left|\sin z\right|^2+ m^{}_3 \left|\cos z\right|^2\right)} \left[ A {\cal F} \left( \frac{M^2_1}{M^2_2} \right)  + B {\cal G} \left( \frac{M^2_1}{M^2_2} \right) \right] \; ,
\label{2.2.3}
\end{eqnarray}
with
\begin{eqnarray}
A & = & \left( m^2_3 \left|U^{}_{\alpha 3}\right|^2 - m^2_2 \left|U^{}_{\alpha 2}\right|^2 \right) {\rm Im}\left(\sin^2 z\right)
\nonumber \\
&& + \sqrt{m^{}_2 m^{}_3} \left[ \left(m^{}_3 - m^{}_2 \right) {\rm Im}\left(U^{}_{\alpha 2} U^*_{\alpha 3}\right) {\rm Re}\left(\cos z \sin z\right) \right.
\nonumber \\
&& + \left. \left(m^{}_2 + m^{}_3 \right) {\rm Re}\left(U^{}_{\alpha 2} U^*_{\alpha 3} \right) {\rm Im}\left(\cos z \sin z\right) \right] \;, \nonumber \\
B & = & m^{}_2 m^{}_3 \left( \left|U^{}_{\alpha 3}\right|^2 - \left|U^{}_{\alpha 2}\right|^2 \right) {\rm Im}\left(\sin^2 z\right)
\nonumber \\
&& + \sqrt{m^{}_2 m^{}_3} \left[ \left(m^{}_3 - m^{}_2 \right) {\rm Im}\left(U^{*}_{\alpha 2} U^{}_{\alpha 3}\right) {\rm Re}\left(\cos z \sin z\right) \right.
\nonumber \\
&& + \left. \left(m^{}_2 + m^{}_3 \right) {\rm Re}\left(U^{*}_{\alpha 2} U^{}_{\alpha 3} \right) {\rm Im}\left(\cos z \sin z\right) \right] \;.
\label{2.2.4}
\end{eqnarray}
Finally, $\kappa$ is the efficiency factor accounting for the washout effects due to the inverse-decay and lepton-number-violating scattering processes, which is dependent on the washout mass parameters $\widetilde m^{}_{i \alpha} = |(M^{}_{\rm D})^{}_{\alpha i}|^2/M^{}_i$. In our numerical calculations, we will employ the following empirical fit formula \cite{giudice} to calculate the values of $\kappa$
\begin{eqnarray}
\frac{1}{\kappa(x)} \simeq \frac{3.3 \times 10^{-3} ~{\rm eV}}{x} + \left( \frac{x} {5.5 \times 10^{-4} ~{\rm eV}} \right)^{1.16} \;.
\label{2.2.5}
\end{eqnarray}

\begin{figure*}
\centering
\includegraphics[width=6in]{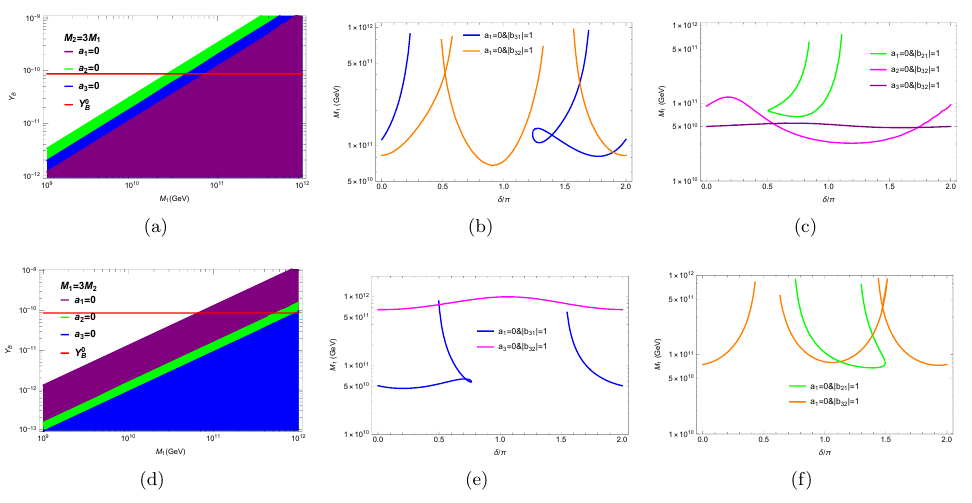}
\caption{ In Scenario A, the allowed ranges of $Y^{}_{\rm B}$ as functions of $M^{}_1$ in the cases of $a^{}_i =0$ for the benchmark values of $M^{}_2 = 3 M^{}_1$ (a) and $M^{}_1 = 3 M^{}_2$ (d). The values of $M^{}_1$ for leptogenesis to be viable as functions of $\delta$ in the cases of $a^{}_i =0$ together with $b^{}_i = b^{}_j$ for $M^{}_2 = 3 M^{}_1$ (b-c) and $M^{}_1 = 3 M^{}_2$ (e-f). }
\label{lepto}
\end{figure*}

Our numerical results for the leptogenesis calculations are shown in Fig.~\ref{lepto}. We first consider the cases of $a^{}_i =0$, where $\delta$ and $\sigma$ are subject to no constraints. In these cases, for the benchmark values of $M^{}_2 = 3 M^{}_1$ (a) and $M^{}_1 = 3 M^{}_2$ (d), the allowed ranges of $Y^{}_{\rm B}$ are shown as functions of $M^{}_1$, obtained by allowing $\delta$ and $\sigma$ to vary freely. From these results one can read in the respective cases the minimally allowed values of $M^{}_1$ for leptogenesis to be viable. We see that all these cases can accommodate a successful leptogenesis for the lighter right-handed neutrino mass $\sim 10^{11}$ GeV. Then, we consider the further imposition of $b^{}_i = b^{}_j$ on the basis of $a^{}_i =0$, which can help us determine $\sigma$ as a function of $\delta$. In these cases, for $M^{}_2 = 3 M^{}_1$ (b-c) and $M^{}_1 = 3 M^{}_2$ (e-f), the values of $M^{}_1$ for leptogenesis to be viable are shown as functions of $\delta$. We see that all these cases can accommodate a successful leptogenesis for some appropriate combinations of $M^{}_1$ and $\delta$. Note that in the case of $a^{}_1 =0$ together with $a^{}_2 = a^{}_3$ where $\delta$ ($\sigma$) is (not) subject to a constraint (see Eq.~(\ref{2.1.5})), the values of $M^{}_1$ for leptogenesis to be viable can be determined as a function of $\sigma$ analogously.

\section{In the scenario of $N^{}_1$ and $N^{}_2$ being nearly degenerate}

In this section, we perform a parallel study for Scenario B (i.e., $M^{}_{\rm R}$ being of the form in Eq.~(\ref{6})). Such a particular form of $M^{}_{\rm R}$ has only one free parameter, enhancing the predictive power of the model.
In section 3.1, we first figure out the phenomenologically viable particular textures of $M^{}_{\rm D}$. As Ref.~\cite{zero} has shown that the two-zero textures of $M^{}_{\rm D}$ have no chance to be consistent with current experimental results, we restrict our analysis to the one-zero textures of $M^{}_{\rm D}$ and further examine if some equalities among the non-vanishing entries can also hold. In section 3.2, the implications of the obtained particular textures of $M^{}_{\rm D}$ for leptogenesis will be investigated.

\subsection{Particular textures of $M^{}_{\rm D}$}

We recall that the Casas-Ibarra parametrization for $M^{}_{\rm D}$ has been formulated in the basis of $M^{}_{\rm R}$ being diagonal.
To accommodate an $M^{}_{\rm R}$ of the form in Eq.~(\ref{6}) into this parametrization, one can go back to the mass basis of right-handed neutrinos via a basis transformation $U^{}_{\rm R}$:
\begin{eqnarray}
U^T_{\rm R} M^{}_{\rm R} U^{}_{\rm R} = D^{}_{\rm R} \; .
\label{3.1.1}
\end{eqnarray}
It is apparent that the two right-handed neutrinos are degenerate in masses (i.e., $D^{}_{\rm R} = M I$ with $I$ being a $2\times 2$ unit matrix) and $U^{}_{\rm R}$ takes a form as
\begin{eqnarray}
U^{}_{\rm R} = \frac{1}{\sqrt 2} \left( \begin{array}{cc} 1 & 1 \cr - 1  & 1 \end{array} \right) P R^\prime \; ,
\label{3.1.2}
\end{eqnarray}
where $P = {\rm diag} ( {\rm i}, 1 )$ serves to ensure the positivity of the right-handed neutrino mass eigenvalues, and $R^\prime$ is an arbitrary orthogonal matrix arising due to the degeneracy between the two right-handed neutrino masses.
Under the above basis transformation, $M^{}_{\rm D}$ becomes $M^{\prime}_{\rm D} = M^{}_{\rm D} U^{}_{\rm R}$. Now that $M^{\prime}_{\rm D}$ can be parameterized in the Casas-Ibarra form as $M^{\prime}_{\rm D} = {\rm i} U D^{1/2}_\nu R D^{1/2}_{\rm R}$, we obtain a modified Casas-Ibarra parametrization for $M^{}_{\rm D}$ as
\begin{eqnarray}
M^{}_{\rm D} = {\rm i} U D^{1/2}_\nu R D^{1/2}_{\rm R} U^\dagger_{\rm R} \; .
\label{3.1.3}
\end{eqnarray}
Because of $D^{}_{\rm R} = MI$, $R^\prime$ can be absorbed via a redefinition of $R$, so one may simply neglect it. To be explicit, the entries of $M^{}_{\rm D}$ read
\begin{eqnarray}
(M^{}_{\rm D})^{}_{\alpha 1} = \frac{M}{\sqrt 2}
\left(\sqrt{m^{}_i} U^{}_{\alpha i} + {\rm i} \sqrt{m^{}_j} U^{}_{\alpha j} \right) e^{-{\rm i}z} \; ,
\nonumber \\
(M^{}_{\rm D})^{}_{\alpha 2} = - \frac{M}{\sqrt 2}
\left(\sqrt{m^{}_i} U^{}_{\alpha i} - {\rm i} \sqrt{m^{}_j} U^{}_{\alpha j} \right) e^{{\rm i}z} \; ,
\label{3.1.4}
\end{eqnarray}
with $i =2$ and $j =3$ ($i=1$ and $j=2$) in the NO (IO) case.

\begin{figure*}
\centering
\includegraphics[width=6in]{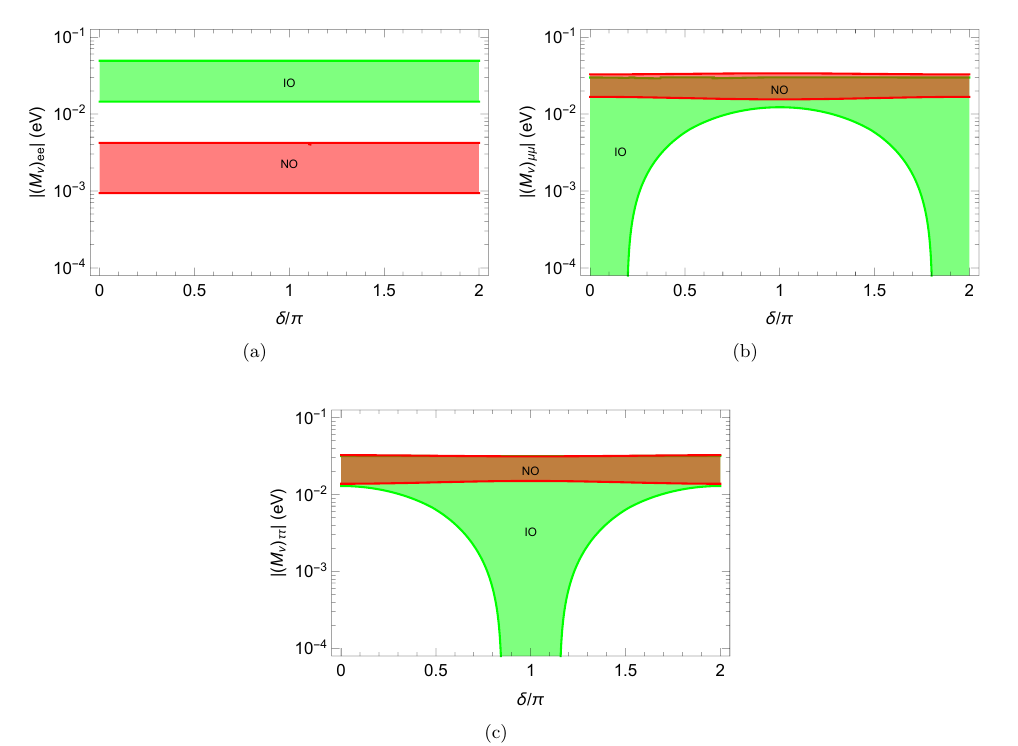}
\caption{ The allowed ranges of $|(M^{}_\nu)^{}_{ee}|$ (a), $|(M^{}_\nu)^{}_{\mu \mu}|$ (b) and $|(M^{}_\nu)^{}_{\tau\tau}|$ (c) as functions of $\delta$ in the NO (red) and IO (green) cases. In obtaining these results, $\sigma$ is allowed to vary in the whole range of $[0, \pi]$, while the other parameters in their $3\sigma$ ranges. }
\label{Mnu}
\end{figure*}

Now we are ready to study the one-zero textures of $M^{}_{\rm D}$. As in Scenario A, we just consider the cases of $(M^{}_{\rm D})^{}_{\alpha 1} =0$. For these cases, one arrives at the following constraint on the low-energy neutrino observables
\begin{eqnarray}
\sqrt{m^{}_i} U^{}_{\alpha i} + {\rm i} \sqrt{m^{}_j} U^{}_{\alpha j} =0 \;,
\label{3.1.5}
\end{eqnarray}
from Eq.~(\ref{3.1.4}), which can be transformed into
\begin{eqnarray}
m^{}_i U^2_{\alpha i} + m^{}_j U^{2}_{\alpha j} = 0 \;.
\label{3.1.6}
\end{eqnarray}
Taking account the reconstruction relation $M^{}_\nu = U D^{}_\nu U^T$, we see that the condition in Eq.~(\ref{3.1.6}) is actually $(M^{}_\nu)^{}_{\alpha \alpha} =0$. This observation can be directly verified by using the seesaw formula: for the form of $M^{}_{\rm R}$ under consideration, an $M^{}_{\rm D}$ with $(M^{}_{\rm D})^{}_{\alpha 1} =0$ does lead to an $M^{}_\nu$ with $(M^{}_\nu)^{}_{\alpha \alpha} =0$.
In Fig.~\ref{Mnu}, we plot the allowed ranges of $|(M^{}_\nu)^{}_{\alpha \alpha}|$ as functions of $\delta$ in the NO and IO cases. One can see that in the NO case none of $(M^{}_\nu)^{}_{\alpha \alpha}$ has chance to vanish. But in the IO case $(M^{}_\nu)^{}_{\mu\mu} =0$ or $(M^{}_\nu)^{}_{\tau\tau} =0$ (correspondingly, $(M^{}_{\rm D})^{}_{\mu 1} =0$ or $(M^{}_{\rm D})^{}_{\tau 1} =0$) can hold within the $3\sigma$ level for $\delta \simeq 0$ or $\pi$. Since the values of $\delta$ for $(M^{}_\nu)^{}_{\mu\mu} =0$ and $(M^{}_\nu)^{}_{\tau\tau} =0$ to hold individually are sharply different, they can not hold simultaneously, verifying the conclusion that the two-zero textures of $M^{}_{\rm D}$ have been ruled out by current experimental results \cite{zero}.

In the IO case, the imposition of $(M^{}_{\rm D})^{}_{\mu 1} =0$ will lead to the following constraint on the low-energy neutrino observables
\begin{eqnarray}
 \sqrt{m^{}_1} \left(-s^{}_{12} - c^{}_{12} \tan \theta^{}_{23} s^{}_{13} e^{{\rm i} \delta} \right)
 + {\rm i} \sqrt{m^{}_2} \left(c^{}_{12} - s^{}_{12}\tan \theta^{}_{23} s^{}_{13} e^{{\rm i} \delta} \right) e^{{\rm i} \sigma} = 0   \;,
\label{3.1.7}
\end{eqnarray}
which enables us to determine $\delta$ and $\sigma$, while the result from the imposition of $(M^{}_{\rm D})^{}_{\tau 1} =0$ can be obtained by making the replacement $\tan \theta^{}_{23} \to - \cot \theta^{}_{23}$. The consequences of this equation for the low-energy neutrino observables (see Table~\ref{prediction2}) are also obtained by minimizing the $\chi^2$ function defined in Eq.~(\ref{2.1.6}). In addition to the aforementioned constraint on $\delta$ (i.e., $\delta \simeq 0$ or $\pi$ in the case of $a^{}_2 =0$ or $a^{}_3 =0$), $\sigma$ is constrained to be around $\pi/2$. Furthermore, $\theta^{}_{23}> 45^\circ$ ($< 45^\circ$) is favored in the case of $a^{}_2 =0$ ($a^{}_3 =0$).
For completeness, an analytical-approximation result for $\delta$ and $\sigma$ is derived from Eq.~(\ref{3.1.7}) as
\begin{eqnarray}
\cos \delta \simeq \frac{1}{2 \tan 2 \theta^{}_{12} s^{}_{13} \tan \theta^{}_{23} }  \; ,
\hspace{1cm}
\sin \sigma \simeq -\frac{1}{ \sin 2\theta^{}_{12} } \;,
\label{3.1.8}
\end{eqnarray}
which can help us understand the above numerical results.

\begin{table}[t]
\centering
\begin{tabular}{c|cccccccc}
\hline\hline
conditions & $\delta/\pi$ & $\sigma/\pi$ & $\Delta m^2_{21}$ & $-\Delta m^2_{31}$  & $s^{2}_{12}$  & $ s^{2}_{23} $ & $s^{2}_{13}$ & $\chi^2_{\rm min}$  \\ \cline{1-9}
$a^{}_2 =0$  & 1.94 &  0.48 & 7.42  & 2.497  & 0.329  & 0.595 & 0.02261 & 10 \\
$a^{}_2 =0$ \& $b^{}_{2} =b^{}_{3}$ & 0.00 &  0.50 & 6.82  & 2.583  & 0.343  & 0.525 & 0.02436 & 51 \\
$a^{}_3 =0$  & 1.07 &  0.52 & 7.38  & 2.499  & 0.343  & 0.456 & 0.02381 & 64 \\
$a^{}_3 =0$ \& $b^{}_{2} =b^{}_{3}$  & 1.07 &  0.48 & 6.82  & 2.557  & 0.343  & 0.475 & 0.02436 & 71 \\
\hline\hline
\end{tabular}
\caption{In Scenario B, the predictions for the low-energy neutrino observables under the various conditions imposed on $M^{}_{\rm D}$. The units of $\Delta m^2_{21}$ and $\Delta m^2_{31}$ are $10^{-5}$ eV$^2$ and $10^{-3}$ eV$^2$, respectively. }
\label{prediction2}
\end{table}

Then, we further examine if some equalities among the non-vanishing entries can hold on the basis of $a^{}_2 =0$ or $a^{}_3 =0$, which can be classified into the following three categories based on the relative positions of the involved entries. (In the present scenario, the right-handed neutrino masses are degenerate, so we will also consider the equalities among the entries residing in different columns.) (1) Equalities among the entries residing in the same column. Thanks to the freedom of the rephasing of three left-handed neutrino fields, in order for $a^{}_{i}=a^{}_{j}$ (or $b^{}_{i}=b^{}_{j}$) to hold, one just needs to have $|a^{}_{i}|=|a^{}_{j}|$ (or $|b^{}_{i}|=|b^{}_{j}|$), whose viabilities have nothing to do with $z$ as can be seen from Eq.~(\ref{3.1.4}).
Our numerical calculations show that only $b^{}_{2} =b^{}_{3}$ can hold on the basis of $a^{}_2 =0$ or $a^{}_3 =0$ within the $3\sigma$ level, and the predictions for the low-energy neutrino observables receive no considerable modifications (see Table~\ref{prediction2}). (2) Equalities among the entries residing in different columns and different rows. Also due to the freedom of the rephasing of three left-handed neutrino fields, in order for $a^{}_i = b^{}_j$ (for $i \neq j$) to hold, one just needs to have $|a^{}_i| = |b^{}_j|$, which are dependent on $z$ in the form of $|e^{-2{\rm i}z}|=|e^{2 {\rm Im}(z)}|$. Therefore, given the values of the low-energy neutrino observables (determined from the condition of $a^{}_2 =0$ or $a^{}_3 = 0$ as above), $|a^{}_i| = |b^{}_j|$ can always be achieved for some appropriate values of ${\rm Im}(z)$ (see Table~\ref{z}). (3) Equalities among the entries residing in the same row. Note that this time the rephasing of three left-handed neutrino fields can not allow us to achieve $a^{}_i = b^{}_i$ from $|a^{}_i| = |b^{}_i|$ any more. Since $a^{}_i/b^{}_i$ are dependent on $z$ in the form of $e^{-2{\rm i}z}$, given the values of the low-energy neutrino observables, $a^{}_i = b^{}_i$ can always be achieved for some appropriate values of ${\rm Re}(z)$ and ${\rm Im}(z)$ (see Table~\ref{z}). Finally, we point out that $b^{}_2 = b^{}_3$ and $a^{}_i = b^{}_j$ (or $a^{}_i = b^{}_i$) can hold simultaneously, since their viabilities rely on different parameters.

\begin{table}[t]
\centering
\begin{tabular}{c|cccccc}
\hline\hline
$a^{}_2 =0$ & $a^{}_{1} =b^{}_{2}$ & $a^{}_{1} =b^{}_{3}$ & $a^{}_{3} =b^{}_{1}$ & $a^{}_{3} =b^{}_{2}$  & $a^{}_{1} =b^{}_{1}$  & $a^{}_{3} =b^{}_{3}$   \\ \cline{1-7}
${\rm Re}(z)$  & $-$  &  $-$  & $-$   & $-$   & 1.5  & $-$0.08  \\
${\rm Im}(z)$  & $-$0.20 & $-$0.13  & $-$0.16   & 0.51  & $-$0.87   & 0.58 \\
\hline \hline
$a^{}_3 =0$ & $a^{}_{1} =b^{}_{2}$ & $a^{}_{1} =b^{}_{3}$ & $a^{}_{2} =b^{}_{1}$ & $a^{}_{2} =b^{}_{3}$  & $a^{}_{1} =b^{}_{1}$  & $a^{}_{2} =b^{}_{2}$   \\ \cline{1-7}
${\rm Re}(z)$  & $-$  &  $-$  & $-$   & $-$   & 1.7  & $-$0.06   \\
${\rm Im}(z)$  & $-$5.0 & 0.16 & 0.20  & $-$0.51   & 0.87  & $-$5.7   \\
\hline\hline
\end{tabular}
\caption{In Scenario B, the predictions for ${\rm Re}(z)$ and ${\rm Im}(z)$ under the various conditions imposed on $M^{}_{\rm D}$. }
\label{z}
\end{table}

To summarize, in the scenario of $M^{}_{\rm R}$ being of the form in Eq.~(\ref{6}), for the IO case, the phenomenologically viable particular textures of $M^{}_{\rm D}$ are as follows
\begin{eqnarray}
&& \left( \begin{array}{cc}  \times & \times \cr
0 & \Box \cr \times & \Box \end{array} \right) \;, \hspace{1cm}
\left( \begin{array}{cc}  \Box &  \times \cr
0 & \Box \cr \times & \times \end{array} \right) \;, \hspace{1cm}
\left( \begin{array}{cc}  \Box &  \times \cr
0 & \times \cr \times & \Box \end{array} \right) \;,  \hspace{1cm}
\left( \begin{array}{cc}  \times &  \Box \cr
0 & \times \cr \Box & \times \end{array} \right) \;, \nonumber \\
&& \left( \begin{array}{cc}  \times &  \times \cr
0 & \Box  \cr \Box & \times \end{array} \right) \;, \hspace{1cm}
\left( \begin{array}{cc}  \Box & \Box \cr
0 & \times \cr  \times & \times \end{array} \right) \;, \hspace{1cm}
\left( \begin{array}{cc}  \times & \times \cr
0  & \times \cr  \Box & \Box \end{array} \right) \;,
\label{3.1.9}
\end{eqnarray}
together with their partners obtained by interchanging the second and third rows. Furthermore, the following more restricted textures of $M^{}_{\rm D}$ can also be consistent current experimental results
\begin{eqnarray}
&& \left( \begin{array}{cc}  \Box & \times \cr
0 & \Box \cr \times & \Box \end{array} \right) \;, \hspace{1cm}
\left( \begin{array}{cc}  \times &  \diamondsuit \cr
0 & \Box \cr \diamondsuit & \Box \end{array} \right) \;, \hspace{1cm}
\left( \begin{array}{cc}  \times & \times \cr
0 & \Box \cr \Box & \Box \end{array} \right) \;, \hspace{1cm}
\left( \begin{array}{cc}  \diamondsuit & \diamondsuit \cr
0 & \Box \cr \times & \Box \end{array} \right) \;,
\label{3.1.10}
\end{eqnarray}
together with their partners obtained by interchanging the second and third rows. Finally, the textures of $M^{}_\nu$ that correspond to these textures of $M^{}_{\rm D}$ are also listed in Table~\ref{texture}.

Before proceeding, we give some discussions about the potential impacts of the renormalization group equation (RGE) evolution effect on the texture zeros and equalities of $M^{}_{\rm D}$. Given an $M^{}_{\rm D}(\Lambda^{}_{\rm SS})$ at the seesaw scale, its counterpart $M^{}_{\rm D}(\Lambda^{}_{\rm EW})$ at the electroweak scale can be obtained as $M^{}_{\rm D}(\Lambda^{}_{\rm EW}) \propto {\rm diag}(1- \Delta^{}_{e}, 1 - \Delta^{}_{\mu}, 1-\Delta^{}_{\tau} ) M^{}_{\rm D}(\Lambda^{}_{\rm SS})$ \cite{IRGE}, with
\begin{eqnarray}
\Delta^{}_{\alpha} \simeq \frac{C}{16 \pi^2}\int_{0 }^{\rm ln(
\Lambda^{}_{SS}/\Lambda^{}_{EW}) } y^2_{\alpha} \ {\rm dt} \;,
\label{3.1.11}
\end{eqnarray}
where $t\equiv {\rm ln} \left( \mu / \mu^{}_0 \right)$ with $\mu$ denoting the renormalization scale, and $C= -3/2$ or 1 in the SM or MSSM. From this expression, one can make the following observations. (1) It is direct to see that neither the texture zeros of $M^{}_{\rm D}$ nor the equalities among the entries in the same row are affected by the RGE evolution effect.
(2) Because of the smallness of $y^{}_e$ and $y^{}_\mu$, $|\Delta^{}_e|$ and $|\Delta^{}_{\mu}|$ are negligibly small. Consequently, the equalities between one entry in the first row and another entry in the second row are not affected by the RGE evolution effect. (3) In the SM, $|\Delta^{}_{\tau}|$ is only $\mathcal O(10^{-5})$. Consequently, the equalities between one entry in the third row and another entry in the first two rows are not affected by the RGE evolution effect either.
(4) In the MSSM, $y^{}_\tau$ can be greatly enhanced by a large $\tan{\beta}$ value. Nevertheless, $|\Delta^{}_{\tau}|$ is still smaller than 0.04 for $\tan{\beta} < 50$ and $\Lambda^{}_{\rm SS} \simeq 10^{13}$ GeV. Consequently, the equalities between one entry in the third row and another entry in the first two rows can be broken at the percent level at most, which will be undermined by the experimental uncertainties of the low-energy neutrino observables themselves. For these reasons, the impacts of the RGE evolution effect on the texture zeros and equalities of $M^{}_{\rm D}$ can be safely neglected. Therefore, although we have examined the viabilities of the particular textures of $M^{}_{\rm D}$ using the values of the low-energy neutrino observables measured at low energies, the conclusions about them will also hold at the seesaw scale.

\subsection{Implications for leptogenesis}

In this subsection, we study the implications of the above obtained particular textures of $M^{}_{\rm D}$ for leptogenesis. In order for leptogenesis to work successfully, the degeneracy between the two right-handed neutrino masses must be broken. Here we consider the contributions from the following two effects: (1) the next-to-leading (NLO) seesaw correction; (2) the renormalization group equation (RGE) evolution effect. As one will see, the mass splitting between the two right-handed neutrinos induced by these effects is extremely small, keeping them nearly degenerate. For nearly degenerate right-handed neutrinos, the CP asymmetries for their decays will receive resonant enhancements, realizing the resonant lactogenesis scenario \cite{resonant}. This scenario allows to lower the leptogenesis scale (approximately the right-handed neutrino masses) down to the TeV scale and thus are quite appealing in phenomenology of particle physics and cosmology \cite{low}:
on the one hand, TeV-scale right-handed neutrinos may potentially manifest themselves at the high-energy colliders. On the other hand, TeV-scale leptogenesis can help us evade the tension between the lower bound $\sim 10^{10}$ GeV of the reheating temperature after inflation required by a successful leptogenesis in the scenario of the right-handed neutrino masses being hierarchical and the upper bound $\sim 10^9$ GeV required by avoiding the overproduction of gravitinos in a supersymmetric extension of the SM \cite{gravitino}.

In the scenario under consideration, both of the right-handed neutrinos will contribute to leptogenesis, and all the three lepton flavors should be treated separately (i.e., the three-flavor regime). Accordingly, the final baryon asymmetry is given by
\begin{eqnarray}
Y^{}_{\rm B} = - c r \sum^{}_{\alpha} \kappa \left( \widetilde m^{}_\alpha\right) \sum^{}_i \varepsilon^{}_{i\alpha}  \;.
\label{3.2.1}
\end{eqnarray}
Here the resonantly enhanced CP asymmetries $\varepsilon^{}_{i\alpha}$ are given by
\begin{eqnarray}
\varepsilon^{}_{i\alpha} = \frac{{\rm Im}\left\{ (M^*_{\rm D})^{}_{\alpha i} (M^{}_{\rm D})^{}_{\alpha j}
\left[ M^{}_j (M^\dagger_{\rm D} M^{}_{\rm D})^{}_{ij} + M^{}_i (M^\dagger_{\rm D} M^{}_{\rm D})^{}_{ji} \right] \right\} }{8\pi (M^\dagger_{\rm D} M^{}_{\rm D})^{}_{ii} v^2} \cdot \frac{M^{}_i \Delta M^2_{ij}}{(\Delta M^2_{ij})^2 + M^2_i \Gamma^2_j} \;,
\label{3.2.2}
\end{eqnarray}
where $\Delta M^2_{ij} \equiv M^2_i - M^2_j$, $\Gamma^{}_j= (M^\dagger_{\rm D} M^{}_{\rm D})^{}_{jj} M^{}_j/(8\pi v^2)$ is the decay rate of $N^{}_j$, and $j \neq i$. On the other hand, the efficiency factor is determined by $\widetilde m^{}_\alpha = \widetilde m^{}_{1 \alpha} + \widetilde m^{}_{2 \alpha}$. In the Casas-Ibarra parametrization, for the IO case, $\varepsilon^{}_{i \alpha}$ and $\widetilde m^{}_\alpha$ are recast as
\begin{eqnarray}
\varepsilon^{}_{i \alpha} & = & \frac{ 1 }{2\pi v^2 \widetilde m^{}_i } \cdot \frac{ M^2_0 \Delta M }{ 4 (\Delta M)^2 + \Gamma^2_j} \left(m^{}_1 - m^{}_2\right) {\rm Re}\left(\cos z \sin^* z\right)
\nonumber \\
&& \times \left[ \left(m^{}_1 \left|U^{}_{\alpha 1}\right|^2 + m^{}_2 \left|U^{}_{\alpha 2}\right|^2\right) {\rm Im}\left(\cos z \sin^* z\right) \right.
\nonumber \\
&& \left. + \sqrt{m^{}_1 m^{}_2} \left(\left|\cos z\right|^2 + \left|\sin z\right|^2\right) {\rm Im}\left(U^{*}_{\alpha 1} U^{}_{\alpha 2}\right) \right] \;,
\nonumber \\
\widetilde m^{}_\alpha & = & \left(m^{}_1 \left|U^{}_{\alpha 1}\right|^2 + m^{}_2 \left|U^{}_{\alpha 2}\right|^2\right) \left(\left|\cos z\right|^2 + \left|\sin z\right|^2\right)
\nonumber \\
&& + 4 \sqrt{m^{}_1 m^{}_2} \ {\rm Im}\left(\cos z \sin^* z\right) {\rm Im}\left(U^{*}_{\alpha 1} U^{}_{\alpha 2}\right) \;,
\label{3.2.3}
\end{eqnarray}
with
\begin{eqnarray}
\widetilde m^{}_1 = m^{}_1 \left|\cos z\right|^2+ m^{}_2 \left|\sin z\right|^2 \;, \hspace{1cm} \widetilde m^{}_2 = m^{}_1 \left|\sin z\right|^2+ m^{}_2 \left|\cos z\right|^2 \;,
\label{3.2.4}
\end{eqnarray}
$\Gamma^{}_j = M^{2}_0 \widetilde m^{}_j/(8\pi v^2)$ and $\Delta M \equiv M^{}_2 -M^{}_1$. Note that we have replaced $M^{}_1$ and $M^{}_2$ with a common $M^{}_0$ when their difference is of no significance. One can see that the dependence of $Y^{}_{\rm B}$ on the right-handed neutrino masses is completely described by the function $f(M^{}_0, \Delta M) = (M^2_0 \Delta M)/[4 (\Delta M)^2 + \Gamma^2_j]$.

\begin{figure*}
\centering
\includegraphics[width=6in]{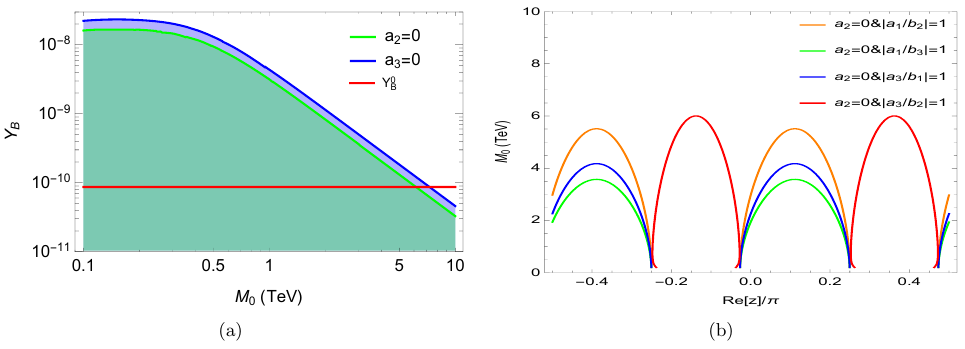}
\caption{ In Scenario B, taking account the contribution of the NLO seesaw correction to the mass splitting between the two right-handed neutrinos, (a) the allowed ranges of $Y^{}_{\rm B}$ as functions of $M^{}_0$ in the cases of $a^{}_2 =0$ and $a^{}_3 =0$; (b) the values of $M^{}_0$ as functions of ${\rm Re}(z)$ for leptogenesis to be viable in the cases of $a^{}_2 =0$ together with $a^{}_i = b^{}_j$.  }
\label{NLO}
\end{figure*}

Let us first consider the contribution of the NLO seesaw correction to the mass splitting between the two right-handed neutrinos: the right-handed neutrino mass matrix becomes
\begin{eqnarray}
M^{\prime}_{\rm R} & = & M^{}_{\rm R} + \frac{1}{2} \left[ \left( M^{}_{\rm D} M^{-1}_{\rm R}  \right)^\dagger M^{}_{\rm D}+ M^{T}_{\rm D} \left( M^{}_{\rm D} M^{-1}_{\rm R}  \right)^* \right] \;,
\label{3.2.5}
\end{eqnarray}
which, in the Casas-Ibarra parametrization, appears as
\begin{eqnarray}
M^{\prime}_{\rm R} = \frac{1}{2} \left( \begin{array}{cc} \widetilde m^{}_2 - \widetilde m^{}_1 -2 {\rm i Re}(\widetilde m^{}_{12})  & 2M + \widetilde m^{}_1 + \widetilde m^{}_2 \cr
2M + \widetilde m^{}_1 + \widetilde m^{}_2 &  \widetilde m^{}_2 - \widetilde m^{}_1 + 2 {\rm i Re}(\widetilde m^{}_{12})  \cr  \end{array} \right) \;,
\label{3.2.6}
\end{eqnarray}
with $\widetilde m^{}_{12} = - m^{}_1 \cos^* z \sin z + m^{}_2 \cos z \sin^* z$.
Given that $M$ is much larger than $m^{}_1$ and $m^{}_2$, to a very good approximation, $\Delta M$ is obtained as
\begin{eqnarray}
\Delta M \simeq  m^{}_2 - m^{}_1  \;.
\label{3.2.7}
\end{eqnarray}
In Fig.~\ref{NLO}(a), for the cases of $a^{}_2 =0$ and $a^{}_3 =0$ where $z$ is subject to no constraints, we plot the allowed ranges of $Y^{}_{\rm B}$ as functions of $M^{}_0$\footnote{The lower boundary of the temperature keeping the sphaleron process efficient is about 100 GeV \cite{review}.} by allowing $z$ to vary freely. It is found that $Y^{}_{\rm B}$ is roughly inversely proportional to $M^2_0$. This is because, for $\Delta M$ in Eq.~(\ref{3.2.7}), one has $\Gamma^{}_j > \Delta M$ and thus $f(M^{}_0, \Delta M) \simeq (M^2_0 \Delta M)/\Gamma^2_j \propto 1/M^2_0$. From the results we see that leptogenesis can be viable for $M^{}_0 \lesssim 7-8$ TeV. Then, we consider the impact of a further imposition of some equality among the non-vanishing entries. (1) For the further imposition of $b^{}_2 = b^{}_3$ which brings no considerable modifications for the predictions for the low-energy neutrino observables, the results are almost the same. (2) For the further imposition of $a^{}_i = b^{}_j$ whose viability fixes ${\rm Im}(z)$ to some specific value, the requirement for a viable leptogenesis will give a determination of $M^{}_0$ as a function of ${\rm Re}(z)$: in Fig.~\ref{NLO}(b), we plot such results for the cases of $a^{}_2=0$ together with $a^{}_i = b^{}_j$, while the results for the cases of $a^{}_3 =0$ together with $a^{}_i = b^{}_j$ are similar and not explicitly shown. (3) For the further imposition of $a^{}_i = b^{}_i$ whose viability fixes ${\rm Re}(z)$ and ${\rm Im}(z)$ to some specific values, the requirement for a viable leptogenesis will give a determination of $M^{}_0$:
for the case of $a^{}_2 =0$ together with $a^{}_1 = b^{}_1$ ($a^{}_3 = b^{}_3$), $M^{}_0$ is determined to be $\sim 2$ ($\sim 5$) TeV. However, the cases of $a^{}_3 =0$ together with $a^{}_i = b^{}_i$ do not admit a viable leptogenesis.

\begin{figure*}
\centering
\includegraphics[width=6in]{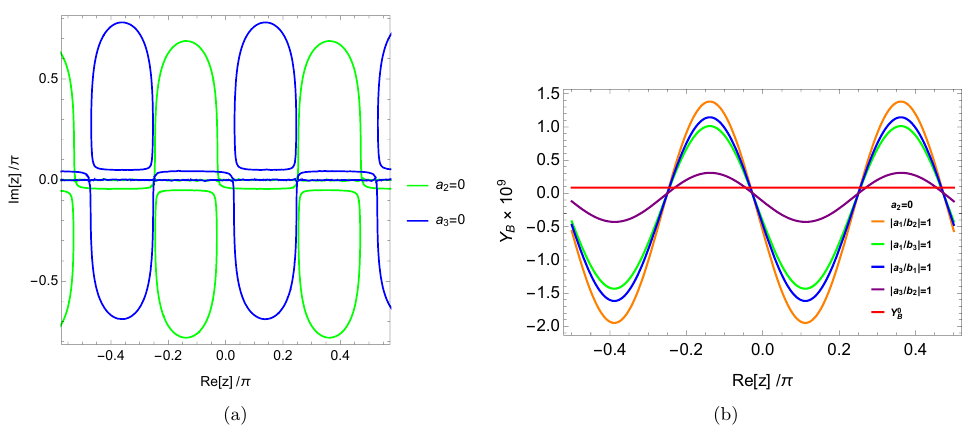}
\caption{ In Scenario B, taking account the contribution of the RGE evolution effect to the mass splitting between the two right-handed neutrinos, (a) the values of ${\rm Im}(z)$ versus ${\rm Re}(z)$ for leptogenesis to be viable in the cases of $a^{}_2=0$ and $a^{}_3=0$; (b) the values of $Y^{}_{\rm B}$ as functions of ${\rm Re}(z)$ in the cases of $a^{}_2 =0$ together with $a^{}_i = b^{}_j$. In obtaining these results, we have taken $M^{}_0 =1$ TeV and $\Lambda = 10^{15}$ GeV as typical inputs. }
\label{RGE}
\end{figure*}

Finally, we consider the contribution of the RGE evolution effect to the mass splitting between the two right-handed neutrinos \cite{RGE1}. This effect will become relevant when the energy scale $\Lambda$ (e.g., the GUT scale) where the right-handed neutrino masses are generated is much higher than the leptogenesis scale $M^{}_0$. At the one-loop level, the RGE evolution behavior of $M^{}_i$ is governed by \cite{RGE2}
\begin{eqnarray}
16 \pi^2 \frac{ {\rm d} M^{}_i}{ {\rm d} t} = \frac{2}{v^2} \left( M^\dagger_{\rm D} M^{}_{\rm D} \right)^{}_{ii} M^{}_i  \;,
\label{3.2.8}
\end{eqnarray}
from which the RGE of $\Delta M$ is immediately obtained as
\begin{eqnarray}
16 \pi^2 \frac{{\rm d} \Delta M }{{\rm d} t} \simeq \frac{2 M^{}_0}{v^2} \left[ \left(  M^\dagger_{\rm D} M^{}_{\rm D} \right)^{}_{22} -  \left(  M^\dagger_{\rm D} M^{}_{\rm D} \right)^{}_{11} \right]  \;.
\label{3.2.9}
\end{eqnarray}
Then, the RGE evolution from $\Lambda$ down to $M^{}_0$ will give a contribution to $\Delta M$ as
\begin{eqnarray}
\Delta M
& \simeq & \frac{M^{2}_0 (m^{}_1 + m^{}_2)}{4 \pi^2 v^2} {\rm Im}\left( \cos z \sin^* z \right) \ln \left(\frac{M^{}_0}{\Lambda} \right) \;.
\label{3.2.10}
\end{eqnarray}
One can see that, taking $M^{}_0 =1$ TeV and $\Lambda = 10^{15}$ GeV as typical inputs, such a contribution to $\Delta M$ is much larger than that from the NLO seesaw correction. It turns out that, except for the logarithmic dependence of $\Delta M$ on $M^{}_0$, both $\Delta M$ and $\Gamma^{}_j$ are proportional to $M^2_0$, leading $f(M^{}_0, \Delta M)$ and thus $Y^{}_{\rm B}$ to be almost independent of $M^{}_0$. For the cases of $a^{}_2=0$ and $a^{}_3=0$, a viable leptogenesis can be achieved for some appropriate values of ${\rm Re}(z)$ and ${\rm Im}(z)$ (see Fig.~\ref{RGE}(a)). Similarly, the further imposition of $b^{}_2 = b^{}_3$ brings no considerable modifications. For the further imposition of $a^{}_i = b^{}_j$, a viable leptogenesis can be achieved for some appropriate values of ${\rm Re}(z)$:
in Fig.~\ref{RGE}(b) we plot the allowed values of $Y^{}_{\rm B}$ as functions of ${\rm Re}(z)$ for the cases of $a^{}_2 =0$ together with $a^{}_i = b^{}_j$ (the results for the cases of $a^{}_3 =0$ together with $a^{}_i = b^{}_j$ are similar and not explicitly shown), from which one can read the value of ${\rm Re}(z)$ for leptogenesis to be viable.
However, the cases of $a^{}_2 =0$ ($a^{}_3 =0$) together with $a^{}_i = b^{}_i$ do not admit a viable leptogenesis.

\section{Summary}

As we know, the seesaw mechanism is the most popular and natural way of generating the light neutrino masses, which also provides an appealing explanation for the baryon asymmetry of the Universe.
Although the information about the low-energy neutrino observables is completely encoded in $M^{}_\nu$, it is still meaningful for us to examine the possible structures of $M^{}_{\rm D}$ and $M^{}_{\rm R}$ for the following two reasons. On the one hand, being more fundamental than $M^{}_\nu$, an investigation on the structures of $M^{}_{\rm D}$ and $M^{}_{\rm R}$ may better help us reveal the possible flavor symmetry underlying the lepton sector as hinted by the particular lepton flavor mixing pattern. On the other hand, when it comes to some high-energy processes such as leptogenesis, the structures of $M^{}_{\rm D}$ and $M^{}_{\rm R}$ will become relevant.
However, the seesaw model consists of much more free parameters than the low-energy neutrino observables, making it difficult to infer the possible structures of $M^{}_{\rm D}$ and $M^{}_{\rm R}$ in light of current experimental results. An attractive way out is to reduce the number of right-handed neutrinos to two, realizing the minimal seesaw model. Nevertheless, the minimal seesaw model still consists of more free parameters than the low-energy neutrino observables. One can further reduce its free parameters by imposing texture zeros (which are usually tied to Abelian flavor symmetries) and equalities (which are usually tied to non-Abelian flavor symmetries) on $M^{}_{\rm D}$ and $M^{}_{\rm R}$.
In this paper, for the minimal seesaw model, following the Occam's razor principle, we explore the particular textures (featuring texture zeros and equalities) of $M^{}_{\rm D}$ in light of current experimental results and leptogenesis, for two particular patterns of $M^{}_{\rm R}$: (A) $M^{}_{\rm R}$ being diagonal ${\rm diag}(M^{}_1, M^{}_2)$; (B) $M^{}_{\rm R}$ being of the form in Eq.~(\ref{6}).

For Scenario A, given that in the NO case the two-zero textures of $M^{}_{\rm D}$ have been ruled out by current experimental results, we (in the NO case, accordingly) turn to the one-zero textures of $M^{}_{\rm D}$, which allow us to fully reconstruct $M^{}_{\rm D}$ only in terms of the low-energy neutrino observables (up to the right-handed neutrino masses). With the help of the reconstruction, one can figure out the ratios among the non-vanishing entries. The results show that the two-zero textures of $M^{}_{\rm D}$ can still hold as a good approximation. In the case of $a^{}_1 =0$, the equalities $a^{}_{2} =a^{}_{3}$, $b^{}_{1} =b^{}_{2}$, $b^{}_{1} =b^{}_{3}$ and $b^{}_{2} =b^{}_{3}$ can hold individually. Furthermore, the equalities $a^{}_{2} =a^{}_{3}$ and $b^{}_{1} =b^{}_{2}$ ($b^{}_{1} =b^{}_{3}$) can hold simultaneously, for which $\sigma$ and $\delta$ are determined to be $0.67 \pi$ and $1.56 \pi$ ($0.35 \pi$ and $1.58 \pi$), respectively. We point out that in the interesting littlest seesaw model \cite{LS} $M^{}_{\rm D}$ just has a texture featuring $a^{}_1 =0$ \& $a^{}_{2} =a^{}_{3}$ \& $b^{}_{1} =b^{}_{3}$. On the other hand,
in the cases of $a^{}_2 =0$ and $a^{}_3 =0$, only the equality $b^{}_2 = b^{}_3$ has chance to hold. And the results in these two cases support that the neutrino sector possesses an approximate $\mu$-$\tau$ flavor symmetry. For leptogenesis, we consider the scenario that there is a hierarchy between $M^{}_1$ and $M^{}_2$. It is found that for both possibilities of $M^{}_1 < M^{}_2$ and $M^{}_2 < M^{}_1$, a successful leptogenesis can be reproduced for the lighter right-handed neutrino mass $\sim 10^{11}$ GeV in all the cases of $a^{}_i =0$ together with $a^{}_i = a^{}_j$ and $b^{}_i = b^{}_j$.

For Scenario B, given that the two-zero textures of $M^{}_{\rm D}$ have been ruled out by current experimental results, we also restrict our analysis to the one-zero textures of $M^{}_{\rm D}$. It is found that in the NO case none of $a^{}_i =0$ can hold while in the IO case $a^{}_2 =0$ and $a^{}_3 =0$ can hold within the $3\sigma$ level. For the IO case, the imposition of $a^{}_2 =0$ ($a^{}_3 =0$), which will lead to $(M^{}_{\nu})^{}_{\mu \mu} =0$ ($(M^{}_{\nu})^{}_{\tau \tau} =0$), restricts $\delta$, $\sigma$ and $s^2_{23}$ to be $1.94 \pi$, $0.48 \pi$ and $0.595$ ($1.07 \pi$, $0.52 \pi$ and $0.456$), respectively. On the basis of $a^{}_2 =0$ ($a^{}_3 =0$), the equalities among the non-vanishing entries can be classified into three categories. (1) For the equalities among the entries residing in the same column, only $b^{}_2 = b^{}_3$ has chance to hold within the $3\sigma$ level. (2) For the equalities among the entries residing in different columns and different rows, $a^{}_i = b^{}_j$ can always be achieved for some appropriate values of ${\rm Im}(z)$. (3) For the equalities among the entries residing in the same row, $a^{}_i = b^{}_i$ can always be achieved for some appropriate values of ${\rm Re}(z)$ and ${\rm Im}(z)$. Furthermore, $b^{}_2 = b^{}_3$ and $a^{}_i = b^{}_j$ (or $a^{}_i = b^{}_i$) can hold simultaneously, since their viabilities rely on different parameters.

In order for leptogenesis to work successfully, the degeneracy between the two right-handed neutrino masses must be broken. This can be achieved by taking account the contributions from the NLO seesaw correction and RGE evolution effect. When the contribution from the NLO seesaw correction is included, $Y^{}_{\rm B}$ will be inversely proportional to $M^2_0$. In the cases of $a^{}_2 =0$ and $a^{}_3 =0$, it has chance to reach the observed value for $M^{}_0 \lesssim 7-8$ TeV. For the further imposition of $a^{}_i = b^{}_j$, the requirement for a viable leptogenesis will give a determination of $M^{}_0$ as a function of ${\rm Re}(z)$. For the further imposition of $a^{}_i = b^{}_i$ in the case of $a^{}_2 =0$, the requirement for a viable leptogenesis will give a determination of $M^{}_0$, while in the case of $a^{}_3 =0$ a viable leptogenesis is not admitted. When the contribution from the RGE evolution effect is taken account, $Y^{}_{\rm B}$ will be dependent on $M^{}_0$ only in a logarithmic manner. In the cases of $a^{}_2 =0$ and $a^{}_3 =0$ (with further imposition of $a^{}_i = b^{}_j$), a viable leptogenesis can be achieved for some appropriate values of ${\rm Re}(z)$ and ${\rm Im}(z)$ (${\rm Re}(z)$). However, as for the further imposition of $a^{}_i = b^{}_i$, a viable leptogenesis is not admitted.

\vspace{0.5cm}

\underline{Acknowledgments} \hspace{0.2cm}
This work is supported in part by the National Natural Science Foundation of China under grant Nos. 11605081 and 12047570, and the Natural Science Foundation of the Liaoning Scientific Committee under grant NO. 2019-ZD-0473.

\end{document}